\documentclass[prd,aps,10pt,twocolumn,nofootinbib,showpacs,notitlepage,tightenlines,longbibliography]{revtex4-2}

\usepackage{amsmath}
\usepackage{amssymb}
\usepackage{dsfont}  
\usepackage{bbm}
\usepackage{graphicx}
\usepackage{soul}
\usepackage{xcolor}

\newcommand{\be}{\begin{equation}}
\newcommand{\ee}{\end{equation}}
\newcommand{\bea}{\begin{eqnarray}}
\newcommand{\eea}{\end{eqnarray}}

\newcommand{\ket}[1]{| #1 \rangle}

\newcommand{\scalar}[2]{\left\langle #1 | #2 \right\rangle}

\newcommand{\ie}{{i.e.}}

\newcommand{\Kc}{\mathcal{K}}
\newcommand{\Hc}{\mathcal{H}}

\begin{document}

\title{Bosonic and fermionic statistics in nonperturbative quantum gravity}
\author{Bekir Bayta\c{s}}
\email{bekirbaytas@iyte.edu.tr}
\affiliation{Department of Mathematics, {\.{I}}zmir Institute of Technology, G{\"{u}}lbah{\c{c}}e, Urla 35430, {\.{I}}zmir, Turkey}
\author{Patrick L. Rodrigues}
\email{patricklrc@ufmg.br}
\author{Nelson Yokomizo}
\email{yokomizo@fisica.ufmg.br}
\affiliation{Departamento de F\'isica - ICEx, Universidade Federal de Minas Gerais, CP 702, 30161-970, Belo Horizonte, MG, Brazil}

\begin{abstract}
The relation between spin and statistics in quantum field theory relies on Poincar\'e invariance, a symmetry that is lost in the presence of a gravitational field, and replaced in general relativity by the principle of general covariance. In a nonperturbative approach to quantum gravity, beyond the picture of gravitational perturbations propagating on a flat background, it is an open question whether the gravitational field must still satisfy a bosonic statistics. By implementing the principle of general covariance through the requirement of invariance under active diffeomorphisms in loop quantum gravity, we find that the space of kinematical states of the gravitational field includes not only bosonic states, but also subspaces of fermionic and mixed statistics.
\end{abstract}
\date{\today}
\maketitle


\section{Introduction}

The spin-statistics theorem in quantum field theory asserts that particles of integer spin are bosons and particles of semi-integer spin are fermions \cite{Pauli:1940zz,Streater:1989vi}. This is a fundamental result in relativistic quantum mechanics that relies on general properties that form the backbone of quantum mechanics in Minkowski spacetime: special relativistic invariance under the Poincar\'e group, the causal structure described by the network of lightcones, boundedness of the energy from below. Such a general framework, designed for the study of relativistic quantum fields in flat spacetimes, cannot be employed in a nonperturbative approach to quantum gravity, however, in a context where there is no fixed background that could provide a determinate causal structure or to which Poincar\'e invariance could refer. Instead, in general relativity, the principle of equivalence extends the range of acceptable coordinate systems from a special family of inertial frames to all smoothly related coordinate systems, and Poincar\'e invariance is replaced by diffeomorphism invariance \cite{Hawking1973,wald1984}. In quantum gravity, in addition, a fixed background of causal relationships may not be available \cite{Hardy:2006uc,Oreshkov:2011er,Chiribella:2009lvz}.

In the canonical formalism for the dynamics of the gravitational field \cite{Arnowitt:1962hi}, diffeomorphism invariance is encoded in a set of constraints. In a nonperturbative approach to canonical quantum gravity based on the Dirac formalism \cite{dirac1967,Hanson:1976cn}, these must be promoted to restrictions in the allowed states of the quantized theory. In this letter, we analyze the consequences of such restrictions to the quantum states of the geometry in the context of loop quantum gravity (LQG) \cite{ashtekar2004background,rovelli2004quantum,thiemann2008modern,Rovelli:2011eq,rovelli2015}. We find that the statistics of local excitations of the gravitational field is related to spin variables describing the state of the geometry in the spin-network representation, but the system is not necessarily bosonic, as one could expect for the gravitational field, usually described by a metric tensor field. Instead, by implementing invariance under active permutations, we show that bosonic, fermionic or mixed statistics can emerge, depending on the graph and spin configuration.


\section{Passive and active transformations in LQG}

We first clarify the distinction between the analogues of passive and active diffeomorphisms in loop quantum gravity. We follow the standard interpretation of the classical theory~\cite{Hawking1973,wald1984}, discussed with great clarity in \cite{Norton1987}, according to which a passive diffeomorphism is a change of coordinates $x^\mu \mapsto x'^{\mu}$, which does not affect the metric tensor field $g$, only its description in terms of components, while an active diffeomorphism is a transformation of the metric tensor field, $g \mapsto \phi^*g$, induced by the action of a diffeomorphism $\phi: M \to M$, where $M$ is the spacetime manifold. In loop quantum gravity, invariance under active spatial diffeomorphisms corresponds to the implementation of the diffeomorphism constraint, as we review below. In addition, we define passive transformations as transformations relating distinct arbitrary labelings of the local regions of the graphs underlying spin-network states, in direct analogy with the arbitrariness of the assignment of coordinates to points of a manifold.

\subsection{Hilbert space of loop quantum gravity} The kinematical Hilbert space $\Kc$ of loop quantum gravity is the space of solutions to the Gauss and diffeomorphism constraints. The Gauss constraint ensures the physical equivalence of distinct choices of local orthonormal frames for the description of the geometry, and the diffeomorphism constraint implements invariance under spatial diffeomorphisms \cite{ashtekar2004background, rovelli2004quantum, thiemann2008modern}. Physical states of the geometry must also satisfy the Hamiltonian constraint that encodes the dynamics of the theory and selects a subspace $\Hc \subset \Kc$. The kinematical structure of LQG is uniquely fixed by the requirement of invariance under the action of spatial diffeomorphisms by the celebrated LOST theorem \cite{Lewandowski:2005jk}, thus providing a solid framework for extracting physical consequences of diffeomorphism invariance.

Throughout this work, we adopt the combinatorial description of the kinematical Hilbert space $\Kc$ of LQG \cite{Rovelli_2011,Rovelli:2011eq}\footnote{Alternatively, the kinematical states can be defined over equivalence classes of embedded graphs $\gamma$ on a fixed three-dimensional manifold $\Sigma$ under extended diffeomorphisms $\phi \in \mathrm{Diff}^*(\Sigma)$ \cite{Fairbairn_2004,Rovelli:2011eq}, where graphs in distinct knot classes define distinct Hilbert subspaces of states of the geometry. In another approach, the group of diffeomorphisms of $\Sigma$ is extended to the automorphism group of the path groupoid of $\Sigma$ \cite{Bahr_2009}, where the states of the geometry do not depend on how the graphs are embedded in $\Sigma$, leading to that obtained through the combinatorial definition.}. Restricting to the subspace $\Kc_\Gamma$ associated with a fixed graph $\Gamma$ amounts to a truncation of the theory to a limited number of degrees of freedom captured by the graph \cite{rovelli2010geometry}, in a coarse-grained representation of the full quantum geometry. We will discuss the implementation of the analogues of passive and active diffeomorphisms in the kinematical Hilbert space $\Kc_\Gamma$ associated with a single abstract graph, in order to focus on the issue of diffeomorphism invariance and disentangle it from questions related to graph refinement or to the dynamics.

Consider a fixed finite abstract graph $\Gamma$, with links $\ell$ oriented from a source node $s(\ell)$ to a target node $t(\ell)$. A configuration of the gravitational field is represented on the graph $\Gamma$ by holonomies $h_\ell$ of the Ashtekar connection integrated along the links $\ell$ of the graph. Let $\Psi(h_\ell)=\scalar{h_\ell}{\Psi}$ be a wavefunction assigning quantum amplitudes to the holonomy configurations $h=\{h_\ell\}$. We denote by $\Hc_\Gamma = L^2[SU(2)^L]$ the space of normalizable wavefunctions with respect to the Haar measure on $SU(2)$. The subspace $\bar{\Kc}_\Gamma \subset \Hc_\Gamma$ selected by the Gauss constraint is formed by $SU(2)$ gauge-invariant states satisfying
\be
\Psi(g_{s(\ell)} h_\ell g_{t(\ell)}^{-1}) = \Psi(h_\ell) \, ,
\ee
where a gauge transformation corresponds to a rotation $g_n \in SU(2)$ at each node $n$. The kinematical Hilbert space $\Kc_\Gamma \subset \bar{\Kc}_\Gamma \subset \Hc_\Gamma$ is the subspace of states that also satisfy the diffeomorphism constraint.

At this level, the residual symmetry imposed by the diffeomorphism constraint is described by graph symmetries \cite{ashtekar2004background,rovelli2004quantum,thiemann2008modern}. In the combinatorial approach, these correspond to graph automorphisms \cite{Rovelli_2011,Rovelli:2011eq}. An automorphism $A: \Gamma \to \Gamma$ is a map of the graph on itself that preserves its connectivity, i.e., such that adjacent nodes are mapped into adjacent nodes. An automorphism acts on $\Hc_\Gamma$ as
\be
(U_A \Psi)(h_\ell) = \Psi(h_{A(\ell)}) \, ,
\label{eq:aut-psi}
\ee
where $A(\ell)$ is the image of the oriented link $\ell$ under the automorphism. The orientation of $A(\ell)$ may not agree with the orientation chosen for the link in the graph, in which case we say there was an inversion. A state is automorphism-invariant if $U_A \ket{\Psi} = \ket{\Psi}$, $\forall A$. 
For details on the action of automorphisms and the implementation of automorphism invariance, we refer to \cite{Baytas:2022mjj,Baytas:2024cux}.

\subsection{Passive transformations} Let us now discuss the implementation of passive transformations. In the classical regime described by general relativity, coordinates $x^\mu$ are introduced for the description of the metric tensor field $g_{\mu \nu}(x)$, but do not possess any physical meaning, being arbitrary labels assigned to points of the spacetime manifold. Similarly, the description of states $\Psi(h_\ell) \in \Hc_\Gamma$ requires choosing a specific presentation of the graph $\Gamma$, with arbitrary labels assigned to its nodes and links. Changing this labeling transforms the explicit form of a wavefunction, as we describe below, but the transformed state must represent the same physical configuration of the geometry, since physical quantities cannot depend on an arbitrary choice of labels. In this work, we identify such relabelings as the analogue of passive diffeomorphisms in general relativity.

For concreteness, let the nodes be labeled by an ordering $n=1,2,\dots, N$, where $N$ is the total number of nodes in the graph, and the links be labeled by ordered pairs of nodes, $\ell=(a,b)$, with $a=s(\ell)$ and $b=t(\ell)$. This provides a specific presentation $\mathcal{P}$ of the graph associated with the given labeling. A Hilbert space $\Hc_{\Gamma,\mathcal{P}}$ is associated with it, formed by the normalizable wavefunctions for the chosen labels. Another presentation $\mathcal{P}'$ could equivalently be used, in which a node originally labeled by $m$ is assigned a new label $m'=\pi(m)$, where $\pi$ is a permutation in the symmetric group $S_N$, and the links are renamed accordingly, $\ell \mapsto \ell'=(\pi(a),\pi(b))$. The normalizable wavefunctions $\Psi'(h_{\ell'})$ referring to the new labels form a Hilbert space $\Hc_{\Gamma,\mathcal{P}'}$. A unitary map
\be
U_\pi: \Hc_{\Gamma,\mathcal{P}} \to \Hc_{\Gamma,\mathcal{P'}}
\ee
then identifies representations of the same physical state in the equivalent presentations of the graph,
\be
\Psi'(h_{\ell'})= [U_\pi(\Psi)](h_{\ell'})=\Psi(h_\ell) \, .
\label{eq:passive-diff}
\ee
The action of such passive transformation on an observable follows from that defined on states in the usual way. If an observation is represented by an observable $\mathcal{O}:\Kc_{\Gamma,\mathcal{P}} \to \Kc_{\Gamma,\mathcal{P}}$ for the labeling $\mathcal{P}$, its possible outcomes and statistical properties are described in a completely equivalent way by the observable
\be
\mathcal{O}' = U_\pi \mathcal{O} U_\pi^{-1}: \, \Kc_{\Gamma,\mathcal{P}'} \to \Kc_{\Gamma,\mathcal{P}'}
\ee
under the alternative labeling $\mathcal{P}'$.

\subsection{Active transformations} Consider now the case of active transformations. In classical gravity, an active diffeomorphism on a spacetime $\langle M,\{U_\alpha,\psi_\alpha\},g\rangle$ is a smooth map of the spacetime manifold into itself, $\varphi: M \to M$, that induces a transformation of the metric,
\be
g \mapsto g'=\varphi^*g \, ,
\label{eq:active-diff}
\ee
without a change of the local charts $\psi_\alpha : U_\alpha \to \mathbb{R}^n$ that assign coordinates $x^\mu$ to the points of the manifold \cite{Norton1987}. In a local neighborhood $U_\alpha$, for a given system of coordinates $\psi_\alpha$, the components of the metric in general change, $g_{\mu \nu}(x) \mapsto g'_{\mu \nu}(x)$. The direct analogue of a diffeomorphism for a graph is a map that preserves its adjacency relations, \ie, an automorphism $A: \Gamma \to \Gamma$. We described the action of an automorphism on quantum states of the geometry in Eq.~\eqref{eq:aut-psi}. We call this action an active transformation in loop quantum gravity.

The nontrivial restriction imposed by the principle of general covariance in general relativity consists in that a configuration of the gravitational field is described by an equivalence class of metric tensor fields $\{\varphi^*g; \varphi \in \mathrm{Diff}(M)\}$ related by the action \eqref{eq:active-diff} of active diffeomorphisms $\varphi$, where $\mathrm{Diff}(M)$ is the diffeomorphism group, and not by a single metric tensor field \cite{Hawking1973,wald1984}. Indeed, in a theory written in the form of tensor equations, so that it is valid for arbitrary systems of coordinates, if diffeomorphic configurations are allowed to be physically distinguishable, then it is possible to obtain an infinite number of physically distinct solutions for a given boundary-value problem, as exemplified by the hole argument of Einstein \cite{Norton1987}. The physical equivalence under active diffeomorphisms must then be incorporated in the framework of any generally covariant theory, and the active diffeomorphisms are gauge transformations associated with the freedom in the choice of coordinates.

Consider now the quantized theory. Let $\ket{\bar{\Psi}} \in \bar{\Kc}_\Gamma$ be an $SU(2)$-invariant state on the graph $\Gamma$. In general, such a state is not in the kinematical Hilbert space $\Kc_\Gamma$, but an automorphism-invariant state $\ket{\Psi}$ can be obtained from it by group averaging,
\be
\ket{\Psi} = \frac{1}{|\mathrm{Aut}(\Gamma)|} \sum_{A \in \mathrm{Aut}(\Gamma)} U_A \ket{\bar{\Psi}} \, ,\label{eq:invariant-projected-state} 
\ee
where $\mathrm{Aut}(\Gamma)$ denotes the automorphism group of $\Gamma$. Since $SU(2)$ invariance is preserved by the procedure, it follows that $\ket{\Psi} \in \Kc_\Gamma$. Conversely, any state in $\Kc_\Gamma$ can be written in the form \eqref{eq:invariant-projected-state}, with $\ket{\bar{\Psi}} \in \bar{\Kc}_\Gamma$. Hence, a generic state $\ket{\Psi} \in \Kc_\Gamma$ is a superposition of all states $U_A \ket{\bar{\Psi}}$ related by the action of the automorphisms, for some $\ket{\bar{\Psi}} \in \bar{\Kc}_\Gamma$. Moreover, if two states $\ket{\bar{\Psi}},\ket{\bar{\Psi}'} \in \bar{\Kc}_\Gamma$ are related by some automorphism $A'$, $\ket{\bar{\Psi}'}= U_{A'} \ket{\bar{\Psi}}$, then they give rise to the same kinematical state, $\ket{\Psi}=\ket{\Psi'}$. Hence, in parallel with the classical theory, a kinematical state of gravity is associated with an equivalence class of states $U_A \Psi$ related by automorphisms, which are the active transformations in LQG. An important difference is that the quantized field is represented by a superposition of such states $U_A \ket{\bar{\Psi}}$, given by Eq.~\eqref{eq:invariant-projected-state}, whereas the classical gravitational field is represented by an equivalence class of diffeomorphic metric tensors $\varphi^*g$.

In summary, for a fixed graph, automorphism invariance constitutes the analogue of invariance under active diffeomorphisms in loop quantum gravity. It imposes a nontrivial restriction on the allowed states of the theory, analogous to the restriction imposed by the principle of general covariance in the continuum. In the next section, we determine the explicit form of these restrictions for several representative graphs.


\section{Spin and statistics}

The restrictions on the allowed states of the geometry imposed by automorphism invariance can be determined from Eq.~\eqref{eq:invariant-projected-state}, with $\ket{\bar{\Psi}} \in \bar{\Kc}_\Gamma$. Any $SU(2)$-invariant state can be expanded in the spin-network basis,
\be
\ket{\bar{\Psi}} = \sum_{j_\ell, i_n} c_{j_\ell, i_n} \, \ket{\Gamma, \{j_\ell\},\{i_n\}} \,,
\label{eq:generic-state}
\ee
where the spin-network states are labeled by spins $j_\ell$ at the links and $SU(2)$ intertwiners $i_n$ at the nodes. The action of an automorphism $A$ on a generic spin-network state was determined in~\cite{Baytas:2022mjj}:
\be
U_A \, \ket{\Gamma, \{j_\ell\},\{i_n\}} = (-1)^R \, \ket{\Gamma, \{j'_\ell\},\{i'_n\}}, 
\label{eq:transformed-spin-network}
\ee
where the configurations of spins and intertwiners are carried around by the
automorphism,
\be
j'_\ell = j_{A^{-1}(\ell)}\, , \quad i'_n = i_{A^{-1}(n)} \,,
\label{eq:new-spins-intertwiners}
\ee
and $(-1)^R$ is a possible sign flip, with $R$ the number of links with semi-integer spins whose image under $A$ has an orientation that disagrees with that of the graph. The transformed state $U_A \ket{\bar{\Psi}}$ can be obtained from Eqs.~\eqref{eq:generic-state}--\eqref{eq:new-spins-intertwiners}, allowing the explicit construction of the group-averaged kinematical state \eqref{eq:invariant-projected-state}.

A spin-network state is a tensor product of intertwiner states at the nodes of the graph,
\be
\ket{\Gamma, \{j_\ell\},\{i_n\}} = \bigotimes_{i=1}^N \ket{i_n} \, .
\ee
Each intertwiner state $\ket{i_n}$ describes a quantum of volume, corresponding to a state of a quantized polyhedron associated with the node \cite{rovelli2004quantum,rovelli2015,Bianchi:2010gc}. The global geometry is the gluing of such local building blocks according to the adjacency relations dictated by the graph. In this picture, the action \eqref{eq:transformed-spin-network} of an automorphism reduces to
\be
\bigotimes_{i=1}^N \ket{i_n} \mapsto (-1)^R \bigotimes_{i=1}^N \ket{i'_n} \, , \quad i'_n = i_{A^{-1}(n)} \, ,
\ee
\ie, a permutation of the node states, accompanied by a factor of $\pm 1$ that depends on the spin configuration and the automorphism. As for many-body states of indistinguishable particles, the basis vectors of the kinematical space are invariant under permutations, up to a phase $\pm 1$, but with two differences: (i) the permutations are restricted to those implemented by automorphisms, and (ii) the states are not necessarily completely symmetric or antisymmetric, it being possible that a sign $+1$ appears for some permutations and a sign flip $-1$ for others. We will see that a fermionic or bosonic statistics naturally emerges in special sectors of interest. Let us analyze some examples.

\paragraph{Dipole graph.} A dipole graph $K_{2,L}$ is a multigraph formed by two nodes $m,n$ connected by $L$ links. We take $L \geq 4$ and orient all links from $m$ to $n$. The automorphism group, of order $2 L!$, is the direct product $S_{2,L} = S_2 \times S_L$, where $S_2$ permutes the nodes and $S_L$ permutes links. It decomposes into two sets $\{(e, \sigma); \sigma \in S_L \}$ and $\{(\pi, \sigma); \sigma \in S_L \}$, where $e$ denotes the identity and $\pi$ is the transposition of the two nodes. Elements of the former (non-inversions) preserve the orientation of all links, and elements of the latter (inversions) reverse their orientations. Consider the Hilbert space $\bar{\mathcal{K}}_{K_{2,L},j_\ell}$ of $SU(2)$-invariant states for a fixed spin configuration $j_\ell$, with $j_\ell \in \mathbb{N}/2$. Due to the Clebsch-Gordan conditions, the number of semi‑integer spins must be even. For automorphisms that are non-inversions, we have $R=0$ for any spin configuration, giving $(-1)^R=+1$. For inversions, $R$ must be even, and again $(-1)^R=+1$ for any spin configuration. Therefore, the action of any automorphism $A \in S_{2,L}$ on a spin-network state $\ket{K_{2,L}, \{j_\ell\},\{i_m, i_n\}} = \ket{i_m}\otimes \ket{i_n}$ reads
\be
U_A (\ket{i_m} \otimes \ket{i_n}) =  \ket{i_{A^{-1}(m)}} \otimes \ket{i_{A^{-1}(n)}} \, ,
\ee
without a sign flip associated with the permutations. Because the transformed state acquires no overall sign change under permutations, the statistics of kinematical states on the dipole graph can be identified as bosonic, and the kinematical Hilbert space $\mathcal{K}_{K_{2,L},j_\ell}$ is spanned by totally symmetric states.

\paragraph{Pentagram.} The pentagram $K_5$ is a complete graph composed of five nodes and ten links, each pair of nodes being connected by a link. We orient each link $(n,n')$ with $n<n'$, that is, from $n$ to $n'$. The automorphism group of $K_5$ is the symmetric group $S_5$ of all node permutations. A generic permutation is a product of elementary transpositions of two nodes. It can be checked that an odd number of links have their orientations reversed for any elementary transposition. We consider the Hilbert space $\bar{\mathcal{K}}_{K_5,j_0}$ of $SU(2)$-invariant states with equal spins $j_\ell=j_0$ at all links. For integer $j_0$, there are no links with semi-integer spins to be inverted, and $R=0$ for all automorphisms. The action on a spin-network state reads
\be
U_A \bigotimes_{n=1}^5 \ket{i_n} = \bigotimes_{n=1}^5 \ket{i_{A^{-1}(n)}} ,  \enspace \text{for } j_0 \in \mathbb{N} \, .
\ee
This is a bosonic sector, as for the dipole graph. However, for semi-integer $j_0$, we find that $(-1)^R=+1$ for even permutations and $(-1)^R=-1$ for odd permutations. The action of an automorphism then reads
\be
U_A \bigotimes_{n=1}^5 \ket{i_n} = (-1)^\pi \bigotimes_{n=1}^5 \ket{i_{A^{-1}(n)}} , \enspace \text{for } j_0 \in \mathbb{N} + \frac{1}{2}\, ,
\ee
where $\pi$ is the parity of the permutation. This is explicitly a fermionic sector, and the kinematical Hilbert space $\mathcal{K}_{K_5,j_0}$ is spanned by totally antisymmetric states with respect to the permutation of node states. An exclusion principle is then observed for the node states, expressing that no two nodes can have the same quantum state. Each node represents a quantum of volume, described by a quantized tetrahedron with faces of the same total area; the exclusion principle requires all local quantum tetrahedra to have distinct shapes. For integer spins, there is no such restriction on the local states of the geometry, but the full kinematical state must be completely symmetric. For example, all nodes can be in the same state, as atoms in a Bose-Einstein condensate, or the global state can be the complete symmetrization of a tensor product of any five intertwiners. For a generic spin configuration $\{j_\ell\}$, the state of the geometry is not completely symmetric or antisymmetric, displaying a mixed symmetry.

\paragraph{Complete graphs.} A complete graph $K_N$ is composed of $N$ nodes and $N(N-1)/2$ links, each pair of nodes being connected by a link. This generalizes the previous case of the pentagram, and similar results follow. The authomorphism group of $K_N$ is $S_N$. It is generated by elementary transpositions of two nodes. Each elementary transposition reverses the orientation of an odd number of links, for any choice of orientations for the links. We consider the Hilbert space $\bar{\mathcal{K}}_{K_N,j_0}$ of $SU(2)$-invariant states with equal spins $j_\ell=j_0$ at all links. For integer $j_0$, the kinematical states are bosonic, while for semi-integer $j_0$, they are fermionic.


\section{Summary and discussion}
\label{sec:discussion}

A spin-network state in loop quantum gravity describes a quantum state of the spatial geometry built from quantized polyhedra glued together according to the combinatorial structure of an associated graph $\Gamma$. Each node of the graph represents a quantum of volume, corresponding to a single quantum polyhedron. A spin-network state based on a graph with $N$ nodes can thus be viewed as an $N$-body quantum system whose constituents are the quantum polyhedra associated with the nodes of the graph. In the combinatorial description of the kinematical Hilbert space of LQG, invariance under active diffeomorphisms translates into invariance under automorphisms for states based on a fixed graph $\Gamma$, which makes nodes related by the action of an automorphism indistinguishable. This leads to strong implications for highly symmetric graphs. For the complete graph $K_N$, in which any pair of nodes is related by an automorphism, automorphism invariance reduces to permutation symmetry, and the polyhedra behave as indistinguishable quantum systems.

We found that, with respect to the exchange of nodes in the complete graph $K_N$, the nonperturbative quantum geometry admits not only bosonic states, as could be intuitively expected, but also fermionic states, as well as sectors of mixed statistics, depending on the configuration of the spins in the spin-network representation. If all links carry the same spin $j_0$, the states are bosonic for integer $j_0$, and fermionic for half-integer $j_0$. A nonuniform distribution of spins allows for states that are neither completely symmetric nor completely antisymmetric, if no additional conditions beyond automorphism invariance are imposed. One may expect further restrictions on the allowed statistics to be enforced by the Hamiltonian constraint, which we do not address here. 

We analyzed the consequences of the diffeomorphism constraint on the kinematical Hilbert space $\Kc_\Gamma$ associated with a fixed abstract graph $\Gamma$ in the combinatorial approach, establishing the existence of sectors of bosonic as well as nonbosonic symmetry under node permutations. It would be interesting to extend this analysis to the full kinematical Hilbert space $\Kc$, which would require, in particular, considering the interplay between graph refinement and graph automorphisms.

In the formulation of the dynamics of loop quantum gravity as a group field theory \cite{Oriti:2017ave}, the local excitations of the geometry are usually assumed to be bosonic. Although nonbosonic statistics have been considered in group field theory models \cite{Gurau_2011}, this has not been done in the context of loop quantum gravity, as noted in \cite{Oriti:2017ave}. Our results provide a motivation for investigating nonbosonic statistics in this setting, including the possibility of mixed statistics depending on the spin configuration. Nonbosonic statistics may also play a role in statistical-mechanical descriptions of the geometry based on ensembles of spin-network states, particularly in regimes far from the classical limit.

Besides deriving the implications of automorphism invariance for the statistics of local excitations of the geometry, we also introduced the analogue in loop quantum gravity of passive diffeomorphisms. Relabelings, which change the arbitrary labels assigned to the nodes and links of a graph, correspond to passive transformations that, similarly to changes of coordinates on a manifold, do not restrict the allowed states of the gravitational field. Instead, they relate equivalent descriptions of the same physical state in distinct representations through a unitary transformation. On the other hand, the analogue of invariance under active diffeomorphisms is automorphism invariance, which selects a proper subspace $\Kc_\Gamma \subset \bar{\Kc}_\Gamma$ of allowed states within the space of $SU(2)$-invariant states. This leads to nontrivial restrictions whose consequences include the exchange statistics of local states of the geometry described in this work.


\begin{acknowledgments}
NY would like to thank Nat\'alia S. M\'oller for insightful discussions. BB acknowledges the support from the Scientific and Technological Research Council of T\"urkiye (T\"UB\.{I}TAK) 3501 under project no.~125F279.
\end{acknowledgments}

\appendix

\end{document}